**Design of efficient vdW thermionic heterostructures from first principles**

By *Xiaoming Wang, Mona Zebarjadi* and *Keivan Esfarjani**

Dr. X. Wang
Institute for Advanced Materials, Devices and Nanotechnology
Rutgers University, Piscataway, New Jersey 08854, United States
Prof. M. Zebarjadi, Prof. K. Esfarjani
Institute for Advanced Materials, Devices and Nanotechnology
Department of Mechanical and Aerospace Engineering
Rutgers University, Piscataway, New Jersey 08854, United States
Email: k1.esfarjani@rutgers.edu

Thermionic transport holds great potential to provide high-efficiency energy conversion devices for waste heat harvesting or cooling.[1-4] Vacuum thermionic refrigerators were proposed by Mahan[1] in 1994 with theoretical efficiency over 80% of the Carnot limit. A current is used to pump electrons from cold cathode to hot anode. To do so, electrons need to overcome the cathode work function barrier, which is on the order of few eVs in typical metals. Therefore, vacuum thermionic refrigerators can only operate at very high temperatures. To overcome this limitation, Shakouri et al.[2] and Mahan et al.[3] proposed solid-state thermionic devices using semiconductor heterostructures. In these structures, a semiconductor acts as a potential barrier for carrier transport, thereby, lowering the work function from few eVs to less than 1 eV. With proper band alignment and doping, there will be asymmetric barriers for electrons and holes. Thus, one type of carriers will dominate the thermionic transport, and contribute to thermionic cooling. However, conventional solid-state thermionic devices usually suffer from the backflow of the heat from the hot side to the cold side due to the significant thermal conductance of the semiconductor layer, as a result, lowering the device performance.

Van der Waals heterostructures[5, 6] based on two dimensional (2D) layered materials are good candidates for the barrier layers in solid-state thermionic devices due to their weak thermal conduction in the cross-plane direction[7] as a result of the weak van der Waals interactions and phonon interface scattering. In addition, the band gap can be easily tuned via changing the number of the 2D material layers, resulting in easy control of the barrier height for thermionic transport. Thus, here, we proposed a thermionic device, which can has good performance, based on graphene/phosphorene/graphene van der Waals heterostructures. Phosphorene, a single layer of black phosphorus that is much like bulk graphite, is another 2D layered material discovered very recently.[8] One of the most intriguing properties of phosphorene is that the band gap can be continuously tuned from 1.5-2.0 eV for a monolayer down to 0.31-0.36 eV in bulk.[8-12] The weak interlayer van der Waals interactions greatly reduce the thermal transport in the cross-plane direction. All these features make phosphorene a potential material with tunable barrier height in design of thermionic refrigerators.

In this work, the thermionic transport properties of graphene/phosphorene/graphene van der Waals heterostructures in contact with gold electrodes (Au/G/P/G/Au) is being studied by using density functional theory (DFT)-based first principles calculations combined with real space Green's function (GF) formalism. We show that for monolayer phosphorene, quantum tunneling dominates the electric current. Varying the number of layers can change the barrier heights and doping characteristics of phosphorene. The cooling coefficient of performance can be enhanced using five layers of phosphorene, due to elimination of the quantum tunneling effect. The phonon thermal conductance of the proposed device is only 4.1 MW/m$^2$K, which is under the lower limit of that for conventional covalently bonded devices. The important features shown in this study are quite fundamental in designing new nanoscale thermionic devices. To our best knowledge, we are the first group to study solid-state thermionic devices by using first principles calculations. This formalism can also be extended to other layered materials.

To model the proposed device, we use open boundary condition along $z$-axis, while periodicity is imposed in the $xy$ plane, as shown in Figure 1a. The heterostructure is composed of G/P/G sandwiched between two semi-infinite gold (111) electrodes. This device configuration can be easily realized in experiments.[13, 14] The graphene layer is used to improve the contact quality and prevent orbital hybridization between gold (111) surface and phosphorene. The stacking arrangements of gold (111) surface, graphene and phosphorene are displayed in Figure 1b,c. The graphene honeycomb lattice can match the gold (111) triangular lattice.[15] The unit cell of G/Au is shown by the red dash rhombus in Figure 1b. In order to construct a cell commensurate with the rectangular lattice of phosphorene shown in Figure 1d, we adopt the green dash square as the unit cell for graphene on gold (111) surface. Finally, the (3×2) phosphorene supercell can match the rectangular graphene lattice well, as shown in Figure 1c. More details about the lattice mismatch and strains can be found in the Supporting Information (SI).

One of the most common failures of DFT is the underestimation of the semiconductor band gap. In the case of phosphorene, the band gap is underestimated by about 40%.[8, 9, 11] The correct examination of the band gap is crucial in establishing the thermionic performance of the device as currents vary exponentially with the barrier height of the layered system. In this study, we make an approximation that only the phosphorene bands in the heterostructure need to be corrected. This simplification is reasonable for two reasons: firstly, normal local functional DFT calculations can describe the band structures of gold and graphene very well; secondly, the non-local correction to the onsite Hamiltonian of phosphorene is much larger than the hopping between phosphorene and graphene ($\mathbf{H}_{PG}$), or gold ($\mathbf{H}_{PAu}$) due to the weak van der Waals interactions. Here, we choose HSE[16] correction for the bands of phosphorene. However, the heavier GW calculations may also be applied to the present treatment.

Figure 2 shows the local band structures of phosphorene calculated by PBE and HSE, respectively. The HSE corrections shift the conduction and valence bands of phosphorene up and down, respectively. As a result, the band gap is enlarged from 0.97 eV to 1.64 eV. Compared to the value of pristine phosphorene, a band gap increase of about 0.1 eV is found when van der Waals interactions are included, see Table 1, consistent with the result from graphene and phosphorene bilayer heterostructure.[6] The increase in the band gap can be attributed to the interaction between *pz* like WFs of graphene and *sp*$^3$ like WFs of phosphorene, as shown in Figure S1a. The band correction can also be identified by the changes of effective masses ($m^*$) at Γ point, as shown in Table 1. Due to HSE corrections, the $m^*$ of both electrons and holes are decreased in ΓX direction while increased in ΓY direction.[11, 17] Moreover, the change of $m^*$ as a result of van der Waals interactions also indicates obvious anisotropy. The $m^*$ change along ΓX direction is more sensitive to the van der Waals interactions.

The work functions of gold and graphene are calculated to be 5.1 eV and 4.6 eV, respectively. A simple Anderson model predicts graphene to be p-doped when it is contact with gold.[15] Note however, that the doping of graphene cannot be predicted by simply aligning the work functions of gold and graphene in their isolated states. The local band structure of graphene is shown in Figure S2. The Dirac point is shifted 0.1 eV below the Fermi level, which means that graphene in the heterostructure, is n-doped.

To investigate the band-bending at the interface, we consider the local density of states (LDOS) of the heterostructure along the direction normal to the interface, as shown in Figure 3a. The white region in the color map indicates the band gap of phosphorene and the Dirac cone of graphene. The symmetric sides with respect to the white area along the horizontal axis indicate the spectrum of gold leads. The highly localized *d* bands and more dispersive *s* bands of gold can be identified by the red and blue regions, respectively. The first layer of gold in contact with graphene has a striking contrast to inner gold layers. This contrast is due to the charge transfer at the interface of gold and graphene. We can also see that there are less charge transfer to the inner gold layers. From a device point of view, the potential barrier height $E_B$ is very important for thermionic transport. The Schottky-Mott (SM) rule provides a simple way to evaluate $E_B$. For *p*-type barrier, $E_B = I - W$, $I$ is the ionization potential of semiconductor and $W$ is the metal work function. The work function of gold (111) surface is calculated to be 5.1 eV and the ionization potentials of phosphorene are found to be 5.2 eV and 5.5 eV for PBE and HSE functionals, respectively. Therefore the SM barrier height for hole transport should be 0.1 eV and 0.4 eV for PBE and HSE calculations, respectively. However, the *p*-type $E_B$ extracted from the DOS of phosphorene shown in Figure 2, are found to be 0.31 eV for PBE and 0.65 eV for HSE. So the SM rule does not make an accurate prediction in the present heterostructure. The failure is attributed to the neglect of interface chemistry in the SM rule, which cannot be omitted here. Band bending arises when two materials with different local charge imbalance are brought in contact. In the present simulation, phosphorene is infinite or periodical in plane with the same periodicity as graphene and gold. In the cross-plane direction, there is only one single layer thickness of phosphorene. So there is no space for band bending within phosphorene or graphene layers. However, by carefully examining the *d* band edges of gold layers along *z* direction, as shown in Figure 3a, we can see that approaching the contacting gold layers from the inner regions, the *d* bands bend up slightly. This is an indication of the charge transfer from gold to graphene.

To investigate the transport properties of the van der Waals heterostructure, we use the real space Green's function method[18, 19] with the tight-binding Hamiltonian constructed by using maximally localized Wannier functions (MLWFs)[20-22]. Figure 4 displays the averaged electron transmission function $\mathcal{T}(E)$ of the Au/G/P/G/Au heterostructure.

$$\mathcal{T}(E) = \sum_{k} \mathcal{T}(E, \mathbf{k}) w_k \qquad (1)$$

where $\mathcal{T}(E, \mathbf{k})$ is the *k* point dependent transmission function in the 2D Brillouin zone perpendicular to the transport direction. $w_k$ is the weight of each transverse *k* point. The blue solid and red dash curves in Figure 4a show the averaged electron transmission function of the heterostructure obtained from PBE and HSE calculations, respectively. HSE corrections widen the gap and lower the transmission, as shown in the inset. Within the gap, the non-zero transmission indicates the existence of quantum tunneling. HSE provides larger potential barrier, therefore, the tunneling current is reduced. For the purpose of displaying the 2D $k$-point dependent transmission function $\mathcal{T}(E,\mathbf{k})$, we use angular averaging and present the data for $\mathcal{T}(E,q)$ where

$$\mathcal{T}(E,q) = \sum_{k} \mathcal{T}(E,\mathbf{k}) \frac{e^{-\frac{\left(q-\sqrt{k_x^2+k_y^2}\right)^2}{2\sigma^2}}}{\sigma\sqrt{2\pi}} \tag{2}$$

$\sigma$ is the parameter controlling the width of the Gaussian distribution. If $q$ is replaced by the coordinates of the $k$ points in the band path for band structure calculations, we can also obtain the band-resolved transmission similarly. Figure 4b,c shows the band-resolved and $q$-resolved transmission, respectively, obtained from HSE calculations. The graphene bands contribute all the transmission within the gap. At the band edges of phosphorene, near the Γ point, the bands have a small contribution to the total transmission, as can be seen from the black curves along ΓX and ΓY directions. In Figure 4c, the cone shape of the $q$-resolved transmission implies that the tunneling are contributed from the $pz$ WFs of graphene since the Dirac cone of graphene is near the X point at $q = \pi/a$ (a is the lattice parameter).

By using the electron transmission function, one can obtain the linear coherent transport of different physical quantities, namely, the electrical conductance $G$, Seebeck coefficient $S$, and electron thermal conductance $\kappa_{el}$, under linear response approximation.[23, 24] The zero bias conductance of the heterostructure at room temperature are 0.0057 $G_0$ ($G_0 = 7.748\times10^{-5}$ Siemens is the quantum of electrical conductance) and 0.0035 $G_0$ from PBE and HSE calculations, respectively. The calculated room temperature Seebeck coefficients are -25.6 μV/K and -29.3 μV/K for PBE and HSE, respectively. The negative $S$ indicates the heterostructure n-doped, which is attributed to the tunneling current from the n-doped graphene layers. Therefore, for monolayer phosphorene, quantum tunneling results in very small Seebeck coefficient and an overall weak thermionic performance.

The band gap and band alignment of phosphorene with respect to gold changes with the number of phosphorene layers.[9] Therefore, we can tune the thermionic barrier height by changing the number of phosphorene layers. Moreover, by adding more layers, we can reduce the quantum tunneling current. To this end, we replace the monolayer phosphorene by bilayer and quintuple layer phosphorene in the heterostructure. We only do PBE calculations for the heterostructure with bilayer (2P) and quintuple layer (5P) phosphorene due to the heavy computational cost. However, we note that our PBE results can represent an estimation of more layers phosphorene due to the layer dependent band alignment[9] and can still provide a guideline for device design. The LDOS of 2P and 5P are displayed in Figure 3b,c. For 2P, the band gap of phosphorene is 0.6 eV, same as that of the isolated configuration. For more than one layer of phosphorene, the van der Waals interaction effect on the band gap is negligible.[6] In the case of bilayer phosphorene, the Fermi level is pinned at the valence band maximum (VBM) of phosphorene, thus there is an Ohmic contact between phosphorene and gold layers. The LDOS of the two phosphorene layers are exactly the same from parity, therefore, still no band bending occurs along the cross-plane direction. On the contrary, there is a clear band bending for five layers (5P). From inner to the outer layers, the bands of phosphorene layers gradually bend up, due to charge transfer from phosphorene to graphene. The estimated band bending from the center to the edge is 0.25 eV. Figure 4 shows the transmission of 2P and 5P structures. In the case of 2P, phosphorene layers are still not thick enough to eliminate the quantum tunneling effect, which results in the non-zero transmission within the band gap, as seen in Figure 4d, similar to that of 1P. We observe no quantum tunneling for 5P, as indicated by the zero transmission right above the Fermi level shown in Figure 4a and the white region in Figure 4e. The room temperature electrical conductance is 0.013 $G_0$ and 0.0011 $G_0$ for 2P and 5P, respectively. The larger value compared to that of 1P is due to the Ohmic contact. The Seebeck coefficient of 2P and 5P is -24 µV/K and 110 µV/K, respectively. The tunneling current from the n-doped graphene layers results in negative Seebeck coefficient of 2P. For 5P, quantum tunneling is removed and the Fermi level is near the valence band, therefore the Seebeck coefficient is positive. The quantum tunneling effect is reduced compared to that of 1P, therefore more heat per charge carrier is transferred across the contacts, which results in the enhancement of the Seebeck coefficient. The thermoelectric parameters of 1P and 5P at typical temperatures, are compared in Table 2.

One of the drawbacks of solid-state thermionic devices compared with vacuum ones is the phonon thermal conductance, which cause the heat flow from hot anode back to cold cathode, thus lowering the device performance. So the phonon thermal conductance is of great importance for solid-state thermionic devices. We adopt a linear chain model[25] and phonon Green's function method[26, 27] to calculate the 1D phonon transmission, which is then extended to 3D under Debye approximation. Figure 5a,b show the phonon transmission function and thermal conductance of the whole heterostructure with 1P, 2P and 5P, respectively. The room temperature phonon thermal conductance of the devices for 1P, 2P and 5P are 6.0 MW/(m$^2$K), 5.7 MW/(m$^2$K) and 4.1 MW/(m$^2$K), respectively. These values are under the lower limit, which is about 10 W/m$^2$K[28], of that for conventional covalently bonded interfaces. Once the phonon thermal conductance $\kappa_{ph}$ is known, we can evaluate the device cooling coefficient of performance (COP):

$$\text{COP} = (J_Q - \kappa_{ph}\Delta T)/(JV) \tag{3}$$

where $J$ and $J_Q$ are the electrical current and electron thermal current, which can be calculated according to the Landauer formular[19]. $V$ is the applied bias voltage, and the second term in the numerator is the phonon thermal backflow current. Figure 5c compares the COP of 1P, 2P and 5P heterostructures with a hot side temperature $T_H$ = 300 K and a temperature difference of $\Delta T$ = 1 K. The COP of 5P is much larger than that of 1P and 2P. From 1P to 5P, the quantum tunneling dominated transport is transformed to thermionic transport resulting in transporting more heat per charge carrier, thus, enhancing the COP. Figure 5d compares the optimized COP with and without considering the phonon thermal conductance at different operating temperatures for 5P. We can see that the phonon thermal conductance has great effect on the performance of the thermionic device. Considering the phonon thermal conductance, the best COP achieved is 18.5 at 600 K corresponding to an equivalent ZT of 0.13, which is significant for nanoscale devices.

Finally, we provide a justification for neglecting the effect of electron-phonon interaction in this work. The study of such effect is by itself the subject of another paper, but here we will attempt to justify its smallness. The electron-phonon interaction inside the phosphorene layer reduces the thermionic performance as part of the heat carried by electrons is absorbed by phonons and result in Joule heating inside the active layer. Roughly half of this heat flows back to the cold side, resulting in lowering of the performance in the case of coolers for instance. To estimate the effect of electron-phonon interactions, we start by estimating the charge carrier drift rate $1/\tau_d$ in phosphorene in the cross-plane direction:

$$\frac{1}{\tau_d} = \frac{v}{d} = \frac{J}{e\rho A d} = \frac{J}{en} \tag{4}$$

where $v$ is the drift velocity, $\rho$ is the carrier charge density, $A$ is the cross section area, $n = \int \mathrm{LDOS}(E) f(E) dE$ is the number of charge carriers in the phosphorene layers. The estimated drift rates at room temperatures for a typical current density of 1 Amp/μm² are about ~$10^{16}$ and ~$10^{15}$ Hz for 1P and 5P with $n$ of 0.00054 and 0.012, respectively. The reported relaxation rate for e-ph coupling in phosphorene from first principles is ~$10^{13}$-$10^{14}$ Hz.[29] Therefore, we see that carriers travel through the phosphorene layer at a rate faster than they can emit or absorb a phonon in that layer. The e-ph contribution in phosphorene is therefore expected to weakly affect the thermionic transport.

In summary, we used first principles DFT calculations, and real space Green's function formalism, to study the thermionic transport across Au/G/P/G/Au van der Waals heterostructures. Our calculations reveal that for monolayer phosphorene, quantum tunneling dominates the transport. By adding more phosphorene layers, one can switch from tunneling dominated transport to thermionic dominated transport, resulting in transporting more heat per charge carrier, thus, enhancing the cooling coefficient of performance. Varying the number of layers of phosphorene, the van der Waals heterostructures can be changed from electron-doped to hole-doped. Band bending can be clearly observed for five layer phosphorene. The phonon thermal conductance of the heterostructure is as small as 4.1 MW/m$^2$K, which is under the lower limit of that for covalently bonded interfaces. We showed electron-phonon interaction inside phosphorene layer is negligible, and estimated the COP of the device including the phonon thermal conductance of the device. Finally, we showed that the proposed Au/G/P/G/Au heterostructure is a promising thermionic device with COP values as large as 18 in the 500-600K range. Further optimization of the barrier height could enhance the COP and equivalent ZT values.

*Computational Methodology*

*DFT calculations details.* To study the structural and electronic properties of the Au/G/P/G/Au van der Waals heterostructure, we use the state-of-the-art DFT based first principles calculations, as implemented in the Quantum ESPRESSO package[30]. We use the generalized gradient approximation for the exchange and correlation functional, as proposed by Perdew-Burk-Ernzerhof (PBE)[31], the GBRV[32] ultrasoft pseudopotential to treat the ion-electron interactions, plane-wave and charge density cutoff energies of 40 and 200 Ry, respectively, and a Monkhorst-Pack[33] $k$ mesh of 4×4×1 to sample the Brillouin zone. To correctly deal with the van der Waals interactions, we employ the dispersion-corrected DFT functional (optB88-vdW)[34, 35] to relax the structure. The relaxed interlayer distances $d_{P/G}$ = 3.43 Å and $d_{Au/G}$ = 3.37 Å are in good agreement with other first principles results.[6, 36] After this relaxation is performed, the rest of the transport calculations are done with the PBE and HSE functionals.

To correct the band structure by HSE, we first construct the full Hamiltonian **H** of the heterostructure based on PBE calculations by using maximally localized Wannier functions (MLWFs)[20-22], as implemented in the WanT package[37] (More details of the Wannier calculations are provided in the SI).

$$\mathbf{H} = \begin{pmatrix} \mathbf{H}_{Au} & \mathbf{H}_{AuG} & 0 & 0 & 0 \\ \mathbf{H}_{GAu} & \mathbf{H}_G & \mathbf{H}_{GP} & 0 & 0 \\ 0 & \mathbf{H}_{PG} & \mathbf{H}_P & \mathbf{H}_{PG} & 0 \\ 0 & 0 & \mathbf{H}_{GP} & \mathbf{H}_G & \mathbf{H}_{GAu} \\ 0 & 0 & 0 & \mathbf{H}_{AuG} & \mathbf{H}_{Au} \end{pmatrix} \quad (5)$$

The diagonal blocks are onsite Hamiltonians for different layers. As mentioned above, we only correct for the diagonal elements representing onsite energy of phosphorene layer. To do so, we perform separate PBE and HSE calculations and obtain the Hamiltonians $\mathbf{H}_P^{PBE}$ and $\mathbf{H}_P^{HSE}$, respectively, using the atomic positons of phosphorene extracted from the heterostructure. The HSE correction term is $\Delta = \mathbf{H}_P^{HSE} - \mathbf{H}_P^{PBE}$. Then, the full Hamiltonian $\mathbf{H}$ can be updated replacing the phosphorene onsite Hamiltonian $\mathbf{H}_P$ by $\mathbf{H}'_P = \mathbf{H}_P + \Delta$. Diagonalizing the Hamiltonian blocks $\mathbf{H}_P$ and $\mathbf{H}'_P$, we can obtain the local band structures of phosphorene.

*Phonon transport calculations.* We use the conservation of the transverse momentum to describe the whole structure as a linear chain for each transverse momentum ($k_x$, $k_y$). Each layer is considered as a single atom interacting with its nearest layer. The chain is extended along the *z* axis of the heterostructure. The nearest neighbor interlayer interaction is calculated from DFT. Since the thermal conductance in the cross-plane of van der Waals heterostructures is mainly contributed from the longitudinal acoustic (LA) phonons[25], the transverse acoustic (TA) and intralayer optical phonons are neglected. Under this approximation, the interatomic force constants become the interlayer force constants, which can be calculated by finite difference method from first principles, taking the whole layer as a rigid group of atoms:

$$\Phi_{ij}^{\alpha} = -\frac{F_{ij}^{u_\alpha} - F_{ij}^{-u_\alpha}}{2u_\alpha} \quad (6)$$

where $i, j$ is the layer index, $\alpha$ is polarization vector, $u_\alpha$ is the rigid displacement of the whole layer, $F$ is the total force on the layer due to the displacement $u_\alpha$.

Once the force constants are known, the Green's function method can be used to calculate the phonon transmission. The procedure is quite similar to that of electron transport. One only needs to substitute $E$ by $\omega^2$ and $\mathbf{H}$ by $\mathbf{\Phi}$ in Eq. (2), where $\omega$ is the phonon frequency. The effective force constants will however depend on the considered transverse momentum. Once the transmission is calculated for a given transverse momentum, the total transmission is obtained by summing the latter over all the k points in the transverse 2-D Brillouin zone:

$$\mathcal{T}(\omega) = \frac{A}{(2\pi)^2} \int \mathcal{T}(\omega, k_x, k_y) dk_x dk_y \qquad (7)$$

In order to get $\mathcal{T}(\omega, k_x, k_y)$, we should know the transverse $k$ dependent force constant $\Phi(k_x, k_y)$. We assume that for interlayer force constant:

$$\Phi_{ij}^\alpha(k_x, k_y) = \Phi_{ij}^\alpha + \sum_{R_x, R_y \neq 0} \Phi_{ij}^\alpha(R_x, R_y) e^{i(k_x R_x + k_y R_y)} \approx \Phi_{ij}^\alpha \qquad (8)$$

where $R_x$ and $R_y$ are real space lattice translation vectors in the $xy$ plane. These translation vectors being large, one can, to a good approximation, neglect the contribution of the above sum. The onsite force constant contains interlayer and intralayer contributions:

$$F_{ii}^\alpha(k_x, k_y) = -\sum_{i \neq j} F_{ij}^\alpha - F_{ii,intra}^\alpha = -\sum_{i \neq j} F_{ij}^\alpha + m_i W_t^2$$

$$W_t^2 = \begin{cases} c_l^2 k_x^2 + c_t^2 k_y^2 & \alpha = x \\ c_t^2 k_x^2 + c_l^2 k_y^2 & \alpha = y \\ c_t^2 (k_x^2 + k_y^2) & \alpha = z \end{cases} \qquad (9)$$

where $m$ is the mass of the layer, $\omega_t$ is the transverse part of phonon frequency, here we use Debye approximation for the phonon dispersion. $c_l$ and $c_t$ are longitudinal and transverse sound velocities, respectively. If we use $q^2 = k_x^2 + k_y^2$, then for $\alpha = z$, $\omega_t^2 = c_t^2 q^2$; for $\alpha = (x, y)$, $\omega_t^2 = \bar{c}^2 q^2$, where

$$\frac{2}{\bar{c}^2} = \frac{1}{c_l^2} + \frac{1}{c_t^2} \qquad (10)$$

In the Debye model, the phonon wave vector cutoff for 2D Brillouin zone is $q_D = \sqrt{4\pi/A}$, $A$ is the cross section area. Then, Eq. (7) can be rewritten as:

$$\mathcal{T}(\omega) = \frac{A}{(2\pi)^2} \int_0^{q_D} \mathcal{T}(\omega, q) 2\pi q \, dq \tag{11}$$

Finally, the phonon thermal conductance can be obtained as:

$$\kappa_{ph} = \frac{1}{2\pi} \int \hbar\omega \, \mathcal{T}(\omega) \frac{\partial n}{\partial T} d\omega \tag{12}$$

where $n(\omega, T) = 1/[\exp(\hbar\omega/k_B T) - 1]$ is Bose-Einstein distribution function, $\hbar$ is the reduced Planck constant, $k_B$ is Boltzmann constant.

In order to validate the above approximations, we calculate the phonon thermal conductance of gold/triple-layer graphene/gold (Au-3G-Au) system. We take the gold sound velocities of $c_l = 3390$ m/s, $c_t = 1290$ m/s. For graphene, $c_{LA} = 21.6$ km/s, $c_{TA} = 13.9$ km/s. For the out of plane ZA mode of graphene, the phonon dispersion is taken as $\omega_t = \beta q^2$, where $\beta = 4.04 \times 10^{-7}$ m$^2$/s which is obtained by fitting the phonon frequency of 14 THz at M point[38].

Figure S3 shows the 1D and 3D phonon transmission of Au-3G-Au system for LA and TA modes, respectively. The phonon thermal conductance of LA mode is 15 MW/(m$^2$K), which is in excellent agreement with fully first principles calculations[39] of 17 MW/(m$^2$K). The value for TA mode is nearly zero. So the phonon thermal conductance contributed from TA mode of this kind of van der Waals structure can be safely neglected.

For phosphorene, it is anisotropic in the *xy* plane. The sound velocities for ZA mode are $c_1 = 1500$ m/s and $c_2 = 2800$ m/s along the ΓX and ΓY directions, respectively.[40] We take the average value of $\bar{c} = \sqrt{c_1 c_2}$ and assume an isotropic dispersion.

*Acknowledgements*
The authors wish to acknowledge SOE HPC cluster of Rutgers and XSEDE, which is supported by National Science Foundation grant number ACI-1053575, for providing the computation resources. The work was supported by National Science Foundation grant number 1403089. We would also like to thank S. B. Cronin for suggesting this problem and useful discussions.

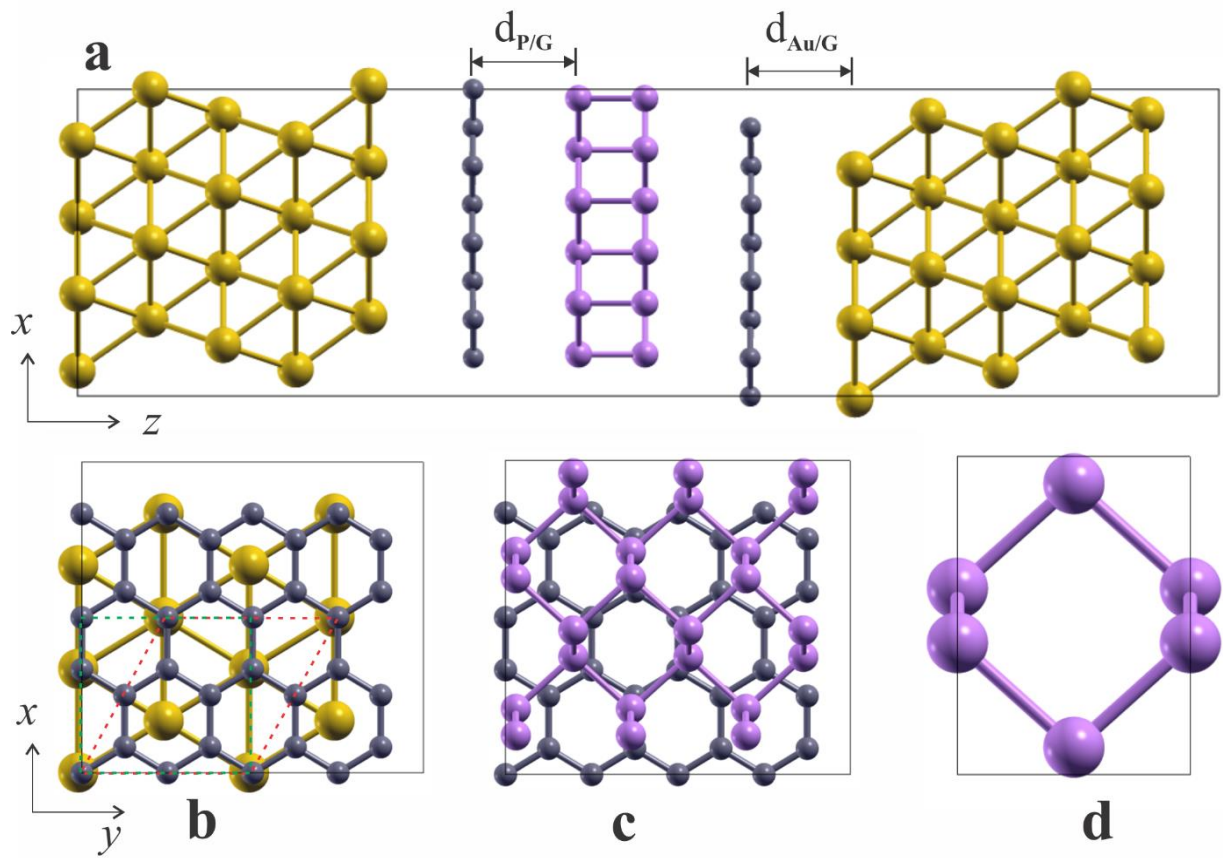

**Figure 1.** (a) Schematic configuration of the Au/G/1P/G/Au heterostructure. Yellow, gray and purple balls denote gold, carbon and phosphorus atoms, respectively. $d_{P/G}$ and $d_{Au/G}$ are the interlayer distances. P, Au and G are short for phosphorene, gold and graphene, respectively. Top-view stacking arrangements of (b) Au/G and (c) G/P in the unit cell used in the simulations. (d) Primitive cell of phosphorene.

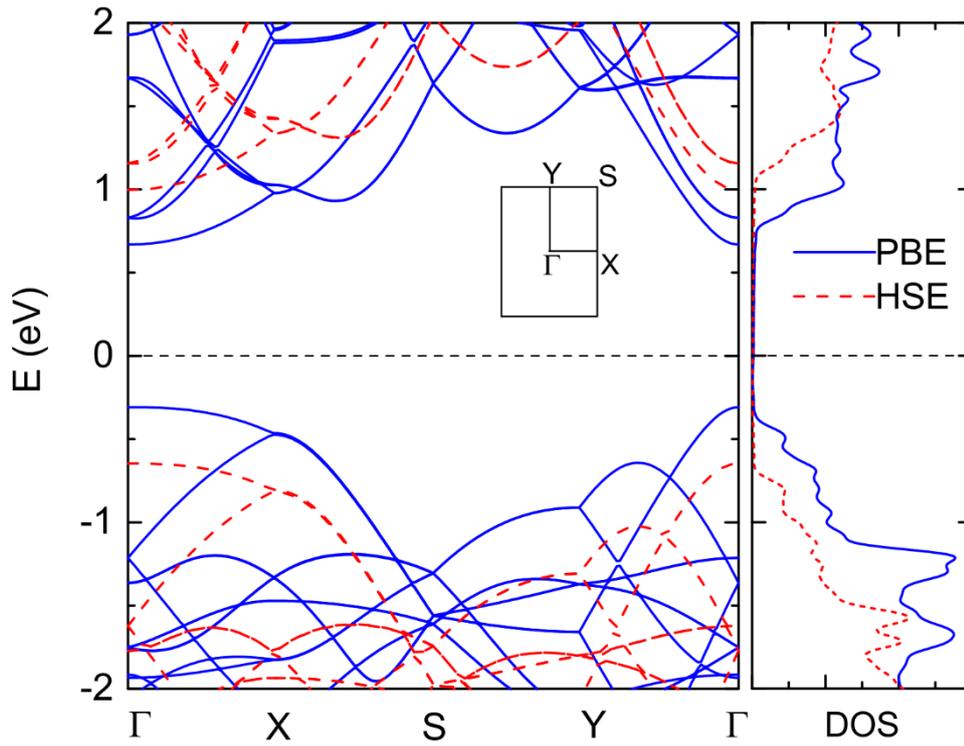

**Figure 2**. Local band structure and density of states (DOS) of 1P phosphorene in the heterostructure calculated by PBE and HSE, respectively. The black dash line indicates the Fermi level.

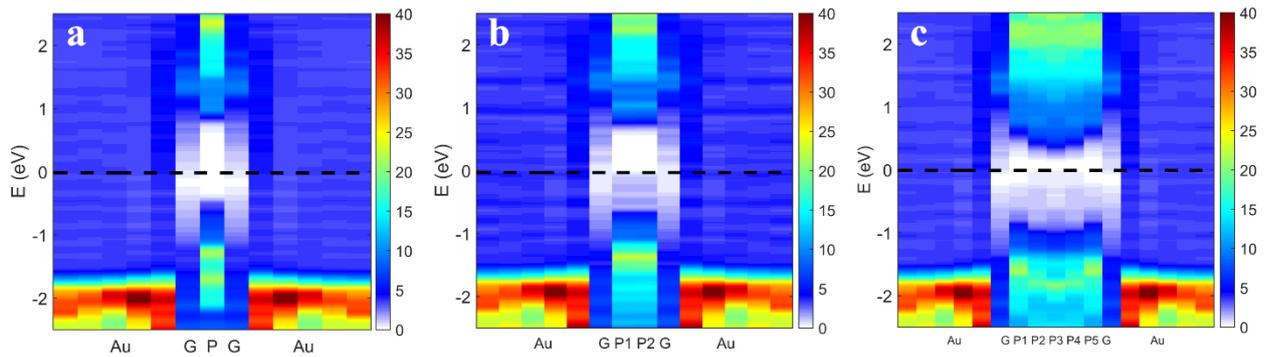

**Figure 3.** Local density of states (LDOS) of the heterostructures for (a) monolayer, (b) bilayer and (c) quintuple layer phosphorene along the direction normal to the interface calculated by PBE. The black dash lines show the Fermi level.

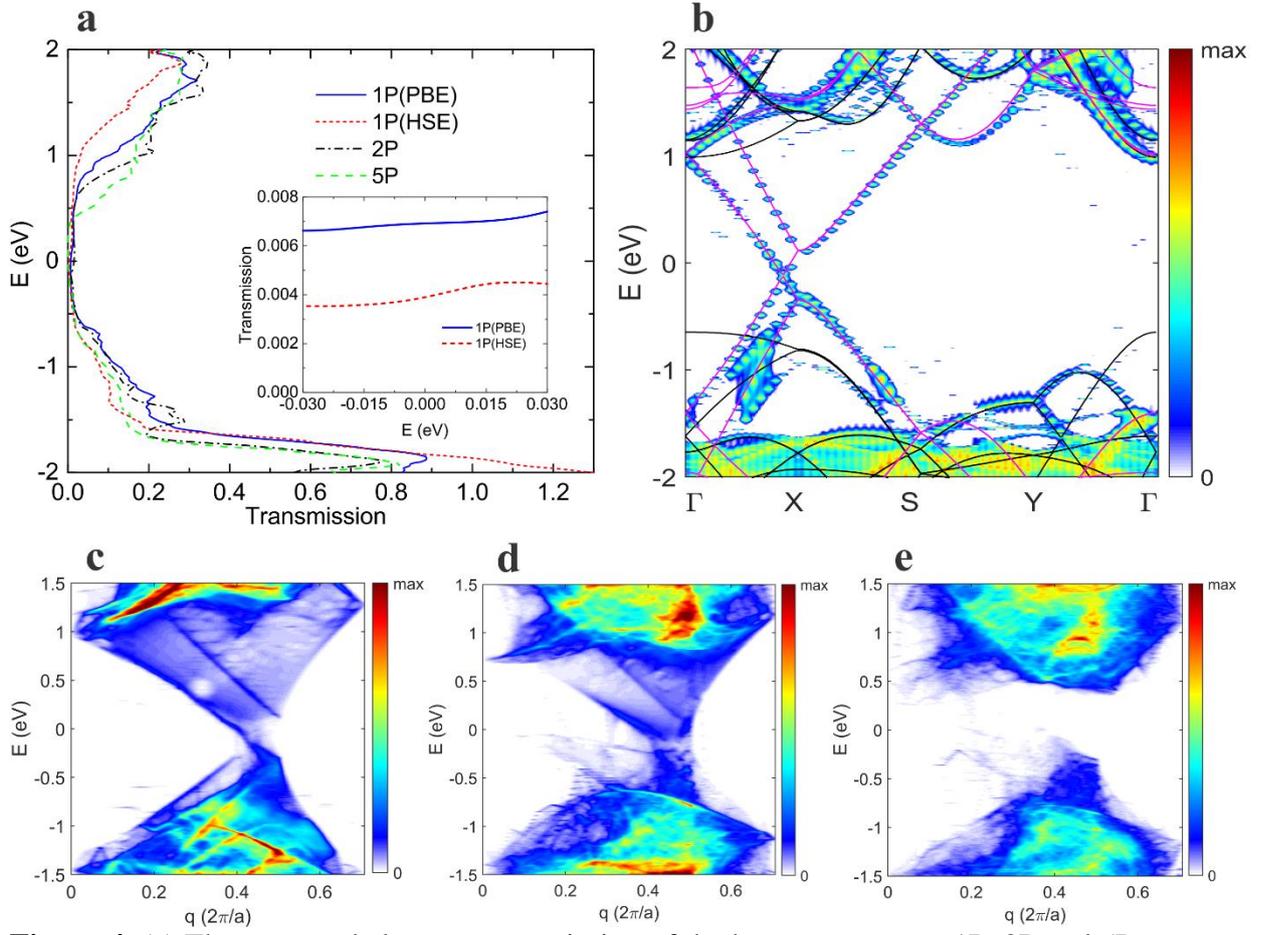

**Figure 4.** (a) The averaged electron transmission of the heterostructures. 1P, 2P and 5P are short for the heterostructures with monolayer, bilayer and quintuple layer phosphorene, respectively. Inset shows the zoom-in of the 1P transmission around Fermi level. (b) The band-resolved transmission of 1P by HSE. The black and magenta curves are the local band structures of phosphorene and graphene, respectively. The 1D $q$-resolved transmission of (c) 1P, (d) 2P and (e) 5P. The data of 2P and 5P are calculated by PBE.

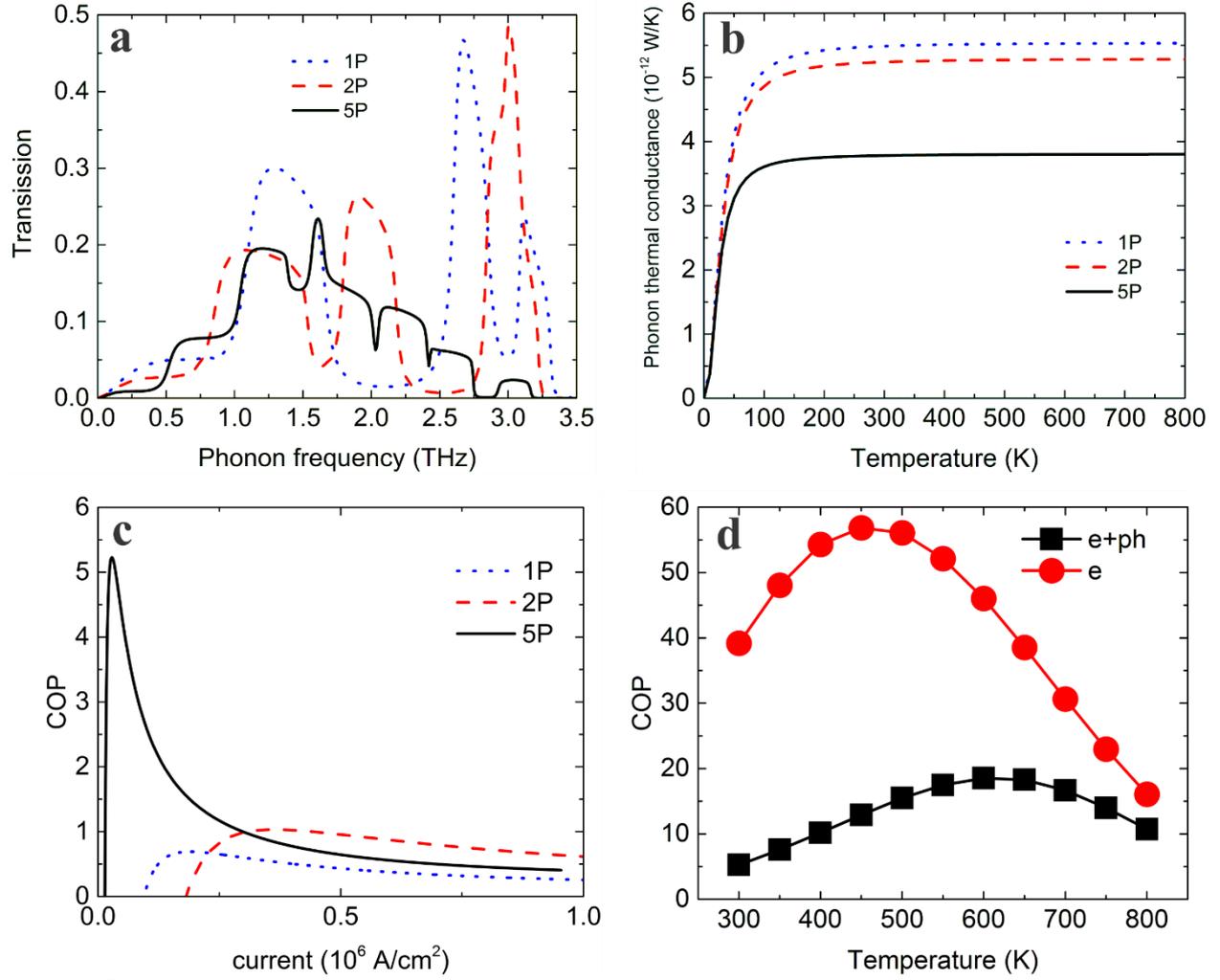

**Figure 5.** (a) Phonon transmission function and (b) thermal conductance of the van der Waals heterostructure with 1P, 2P and 5P. (c) Coefficient of performance (COP) of 1P (HSE data), 2P and 5P with a hot side temperature $T_H = 300$ K and a temperature difference of $\Delta T = 1$ K. (d) Optimized COP of 5P at different operating temperatures with (black squares) and without (red circles) including the phonon thermal conductance.

**Table 1.** Band gap and effective masses of isolated phosphorene and that in the heterostructure.

| Phosphorene | Functional | Band gap (eV) | Carrier | $m^*_{yy}/m_0$ | $m^*_{xx}/m_0$ | |
|---|---|---|---|---|---|---|
| Isolated | PBE | 0.86 | e | 0.16 | 1.24 | |
| | | | | 0.18 | 1.23 | Ref[17] |
| | | | h | 0.13 | 7.14 | |
| | | | | 0.13 | - | Ref[17] |
| | HSE | 1.51 | e | 0.21 | 1.16 | |
| | | | | 0.17 | 1.12 | Ref[11] |
| | | | h | 0.17 | 5.58 | |
| | | | | 0.15 | 6.35 | Ref[11] |
| In the heterostructure | PBE | 0.97 | e | 0.14 | 1.38 | |
| | | | h | 0.14 | 7.22 | |
| | HSE | 1.64 | e | 0.18 | 1.25 | |
| | | | h | 0.17 | 5.49 | |

**Table 2.** Thermoelectric parameters of heterostructures for 1P (HSE data) and 5P at temperatures 200, 300 and 500K. $G$, $\kappa_{el}$, $\kappa_{ph}$ and $S$ are electrical conductance, electron thermal conductance, phonon thermal conductance and Seebeck coefficient, respectively. $GTL$ is the Wiedemann-Franz formula for the electronic thermal conductivity. $L$ is the Lorenz number. $G_0 = 2e^2/h$ and $\kappa_0 = \pi^2 k_B^2 T/(3h)$ are quantum of electrical conductance and thermal conductance, respectively. $e$ is electron charge, $h$ is Planck constant, $k_B$ is Boltzmann constant. The thermoelectric figure of merit $ZT = GS^2T/(\kappa_{el}+\kappa_{ph})$.

| parameter | unit | 1P | | | 5P | | |
|---|---|---|---|---|---|---|---|
| $T$ | K | 200 | 300 | 500 | 200 | 300 | 500 |
| $G$ | $G_0$ | 0.0035 | 0.0035 | 0.0037 | 0.0010 | 0.0011 | 0.0012 |
| $\kappa_{el}$ | pW/K | 1.32 | 2.17 | 4.13 | 0.32 | 0.44 | 1.14 |
| | $GTL$ | 1.00 | 1.09 | 1.18 | 0.85 | 0.71 | 1.01 |
| | $\kappa_0$ | 0.0070 | 0.0076 | 0.0087 | 0.0017 | 0.0016 | 0.0024 |
| $\kappa_{ph}$ | pW/K | 5.42 | 5.49 | 5.52 | 3.75 | 3.78 | 3.80 |
| | $\kappa_0$ | 0.029 | 0.019 | 0.012 | 0.020 | 0.013 | 0.008 |
| $S$ | μV/K | -4.9 | -29.3 | -53.8 | 89.3 | 109.7 | 118.0 |
| | $k_B/e$ | -0.057 | -0.340 | -0.624 | 1.035 | 1.272 | 1.368 |
| ZT | | $2\times10^{-4}$ | 0.01 | 0.04 | 0.03 | 0.07 | 0.13 |

Supporting Information

**Structure relaxation and lattice mismatch**

As the band structure of phosphorene is very sensitive to strain[1], during the geometry optimization, we fixed the lattice parameters of phosphorene to (a = 10.02 Å, b = 9.18 Å). Finally, all the forces on each atom are relaxed within 0.01 eV/Å. After relaxation, the *xy* plane of gold (111) layers have strains of 1.5% and 4.3% along *x*- and *y*-axis, respectively. The graphene layer also has biaxial strains of 1.6% and 7.5% along two axes. The work functions of the gold (111) surface and graphene with the configuration in the heterostructure are calculated to be 5.1 eV and 4.6 eV, respectively, in good agreement with experiments. So we believe that the relaxed structure is reasonable when comparing with experiments.

**Wannier function calculations**

Seven Wannier functions (WFs) with 5 atomic centered *d* orbitals and 2 $\sigma$ orbitals centered on tetrahedron-interstitials (*t* orbital)[2] are used to represent the gold atoms at the interface with graphene, while one *t* orbital is omitted for the inner gold atoms. Four atomic centered *sp*$^3$ orbitals are used for constructing the WFs of each phosphorus atom. For carbon atoms, there are two kinds of WFs with $\sigma$ orbitals centered on mid-bond points and atomic centered *pz* orbitals. After wannierization, the spatial spread of each WF is reduced to 15 bohr$^2$.

**Electron transport.**

To investigate the transport properties of the heterostructure, we use the real space Green's function method with the tight-binding Hamiltonian constructed by using MLWFs. The retarded Green's function of the heterostucture is:

$$\mathbf{G}^r = \left[ (E+i\eta)\mathbf{I} - \mathbf{H} - \mathbf{\Sigma}_L - \mathbf{\Sigma}_R \right]^{-1} \quad (1)$$

where **I** the identity matrix, *E* the electron energy, *iη* is a small imaginary part used to impose causality. $\Sigma_\beta = \mathbf{H}_{C\beta} g_\beta \mathbf{H}_{\beta C}$ $(\beta = L, R)$ denotes the self-energy of the left (L) and right (R) leads, and $\mathbf{g}_\beta$ is the surface Green's function of the leads, which can be calculated from the Hamiltonian matrix elements via an iterative procedure.[3] The electron transmission function $\mathcal{T}(E)$ can be calculated as:

$$\mathcal{T}(E) = \text{Tr}\left(\mathbf{G}^r \mathbf{\Gamma}_L \mathbf{G}^a \mathbf{\Gamma}_R\right) \tag{2}$$

where $\mathbf{G}^a = \left(\mathbf{G}^r\right)^\dagger$ is the advanced Green's function and $\mathbf{\Gamma}_\beta = i\left(\mathbf{\Sigma}_\beta - \mathbf{\Sigma}_\beta^\dagger\right) (\beta = L, R)$. It should be noted that the above method is used for 1D transport. For the case of 3D transport, one should sample the 2D Brillouin zone perpendicular to the transport direction using a fine enough $k$ mesh. For each $k$ point, the $k$ dependent transmission $\mathcal{T}(E, \mathbf{k})$ should be calculated and the averaged transmission would be $\mathcal{T}(E) = \sum_k \mathcal{T}(E, \mathbf{k}) w_k$, where $w_k$ is the weight of each transverse $k$ point. In the transmission calculation, we use a 30×30 $k$ mesh to initialize the k-resolution and further interpolate[4] to a 180×180 $k$ grid to obtain a smooth transmission function.

By using the electron transmission function, one can obtain the linear coherent transport of different physical quantities under linear response approximation[5, 6]:

$$\begin{aligned} G &= -\frac{eJ}{\Delta\mu}\bigg|_{\Delta T=0} = e^2 L_0 \\ S &= -\frac{\Delta\mu}{e\Delta T}\bigg|_{J=0} = L_1 / (eTL_0) \\ \kappa_{el} &= -\frac{J_Q}{\Delta T}\bigg|_{I=0} = \left(L_2 - L_1^2 / L_0\right)/T \\ L_n &= 2/h \int dE\, \mathcal{T}(E) \times (E-\mu)^n \times (-\partial f / \partial E) \end{aligned} \tag{3}$$

where $G$, $S$, and $\kappa_{el}$ denote the electron conductance, Seebeck coefficient, and electron thermal conductance, respectively. $e$ is the charge of electron, and $h$ is the Planck constant, $\mu$ is the chemical potential, $T$ is the absolute temperature, and $f(E, \mu, T) = 1/\left(\exp\left[(E-\mu)/k_B T\right] + 1\right)$ is the Fermi-Dirac distribution function. Based on the electron transmission function, the 2-probe electrical current and electron thermal current can be calculated from the knowledge of the temperature and chemical potentials on both sides (L, R) as:

$$J = 2e/h \int dE\, \mathcal{T}(E)(f_L - f_R) \tag{4}$$

$$J_Q = 2/h \int dE\, \mathcal{T}(E)(E-\mu)(f_L - f_R) \tag{5}$$

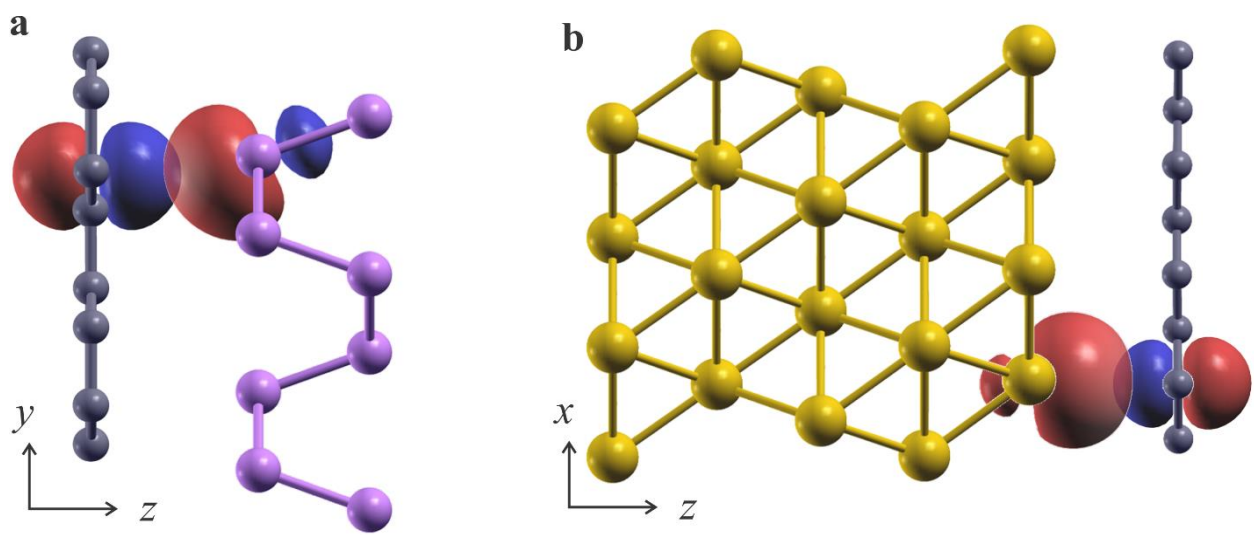

**Figure S1**. (a) pz Wannier function (WF) of graphene overlaps with sp3 WF of phosphorene.

(b) Tetrahedron-interstitial t WF of gold overlaps with pz WF of graphene.

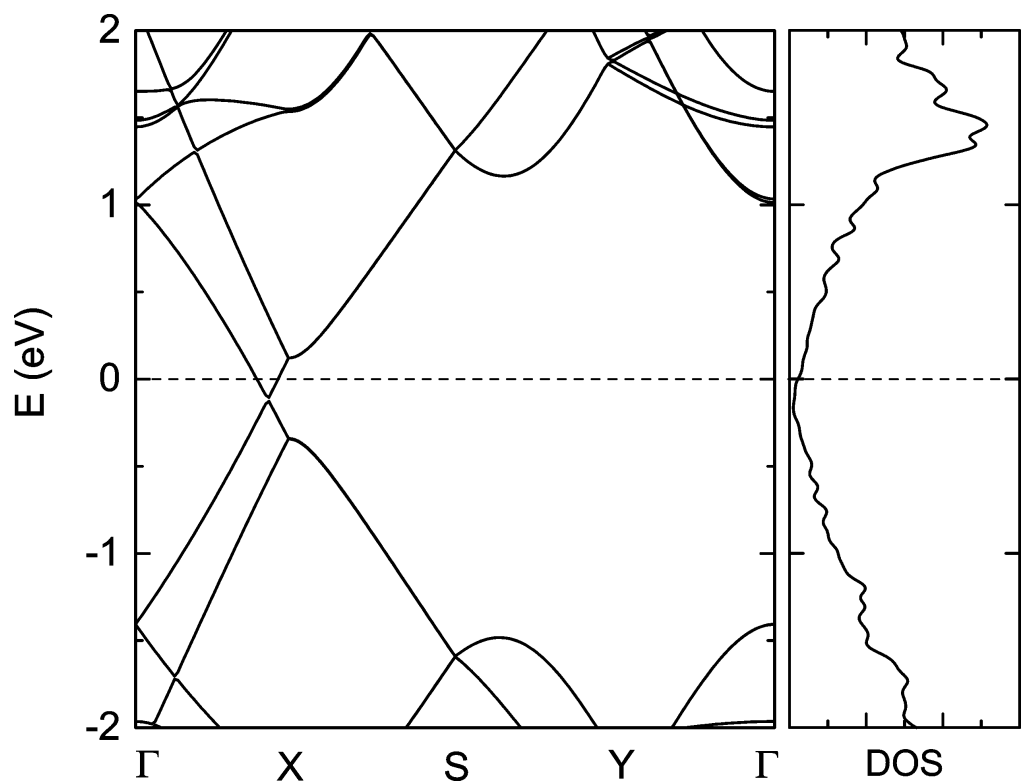

**Figure S2**. Local band structure and density of states (DOS) of graphene in the heterostructure. The black dashed lines indicate the Fermi level.

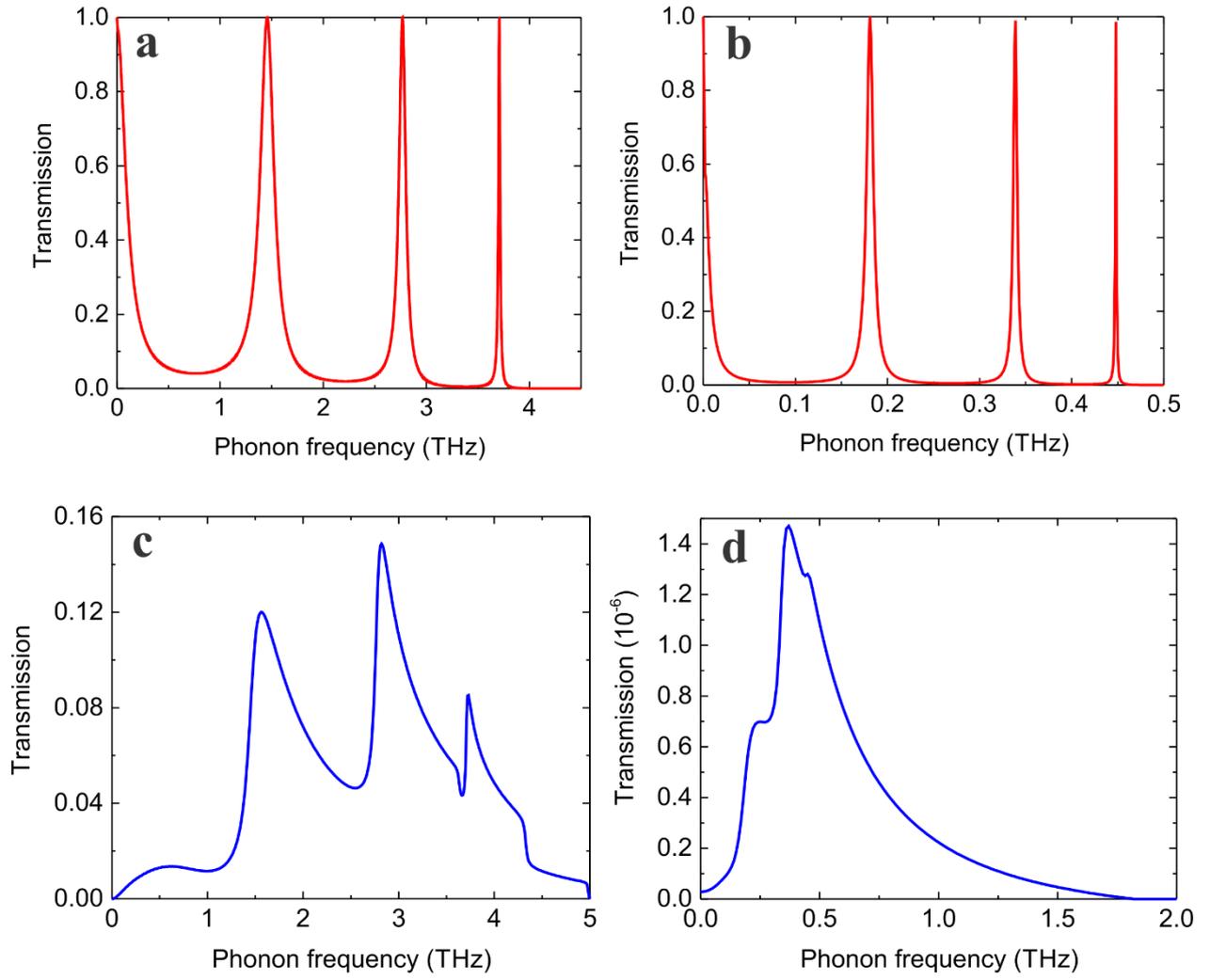

**Figure S3**. Phonon transmission function of Au-3G-Au system. (a) LA in 1D. (b) TA in 1D. (c) LA in 3D. (d) TA in 3D. Note the transmission of the transverse modes is 5 orders of magnitude smaller than the longitudinal one.